\documentclass[oneside,a4paper,12pt]{article}
\usepackage{latexsym}
\usepackage{euscript}
\usepackage{epsfig}
 \large\normalsize
\newcommand{\be}{\begin{equation}}
\newcommand{\ee}{\end{equation}}

\begin{document}
\begin{flushright}
BA-99-82, FTUAM-99-42,\\ IC/99/187, UT-STPD-1/00.
\end{flushright}
\vskip 0.3cm
\begin{center}
{\Large \bf Inflation and Monopoles in Supersymmetric\\
\vskip 0.15cm
  ${\bf SU(4)_c \times
SU(2)_L \times SU(2)_R}$}
\vskip 0.35cm
{ R. Jeannerot$^{1}$, S. Khalil$^{2,3}$, G. Lazarides$^{4}$ and
Q. Shafi$^5$}
\\\vspace*{0.2cm} \small{\textit{$^1$The Abdus Salam International Center
for Theoretical Physics, P.O. Box 586,\\ 34100 Trieste,
Italy.}}\\ \vspace*{1mm}
\small{\textit{$^2$Departmento de Fisica
Te\'orica, C.XI, Universidad Aut\'onoma de Madrid,\\ 28049
Cantoblanco, Madrid, Spain.}} \\
\vspace*{1mm} \small{\textit{$^3$Ain Shams University, Faculty of
Science, Cairo 11566, Egypt.}} \\\vspace*{1mm}
\small{\textit{$^4$Physics Division, School of Technology, Aristotle
University of Thessaloniki,\\ Thessaloniki 540 06, Greece.}}\\
\vspace*{1mm} \small{\textit{$^5$Bartol Research Institute, University
of Delaware Newark, DE 19716, USA.}} \\
\vspace*{0.5cm}

\begin{abstract}
{We show how hybrid inflation can be successfully realized in a 
supersymmetric model with gauge group $G_{PS}=SU(4)_c \times SU(2)_L
\times SU(2)_R$. By including a non-renormalizable superpotential term,
we generate an inflationary valley along which $G_{PS}$ 
is broken to the standard model gauge group. Thus, catastrophic 
production of the doubly charged magnetic monopoles, which are 
predicted by the model, cannot occur at the end of inflation. The 
results of the cosmic background explorer can be reproduced with 
natural values (of order $10^{-3}$) of the relevant coupling constant, and 
symmetry breaking scale of $G_{PS}$ of the order of $10^{16}$ GeV.
The spectral index of density perturbations lies between unity and $0.9$.
Moreover, the $\mu$-term is generated via a Peccei-Quinn symmetry and
proton is practically stable. 
Baryogenesis in the universe takes place via leptogenesis. The low 
deuterium abundance constraint on the baryon asymmetry, the gravitino 
limit on the reheat temperature and the requirement of almost maximal 
$\nu_{\mu}-\nu_{\tau}$ mixing from SuperKamiokande can be simultaneously 
met with $m_{\nu_{\mu}}$, $m_{\nu_{\tau}}$ and heaviest Dirac neutrino
mass determined from the large angle MSW resolution of the solar neutrino
problem, the SuperKamiokande results and $SU(4)_c$ symmetry respectively.}
\end{abstract}

\setcounter{page}{1}
\end{center}

\thispagestyle{empty}

\newpage

\section{ Introduction}

After the recent discovery of neutrino oscillations by the
SuperKamiokande experiment~\cite{japan}, supersymmetric (SUSY) models 
with left-right symmetric gauge groups have attracted a great deal of
attention. These models provide a natural framework for implementing
the seesaw mechanism~\cite{see-saw} which explains the existence of the
small neutrino masses. The implications of these models have been
considered in Ref.\cite{mohapatra}, in the case of the gauge group
$G_{LR}=SU(3)_c \times SU(2)_L \times SU(2)_R \times U(1)_{B-L}$, 
and in Ref.\cite{shafi1} for the SUSY Pati-Salam (PS)~\cite{pati} model
based on the gauge group $G_{PS}=SU(4)_c \times SU(2)_L \times SU(2)_R$.
It was shown that they lead to a constraint version of the minimal
supersymmetric standard model (MSSM).
\vskip 0.25cm
Recently, it was demonstrated~\cite{tye} that the N=1 SUSY PS model 
can emerge as the effective four dimensional limit of brane models 
from type I string theory. This result provides further motivation for 
considering the phenomenological and cosmological implications of this 
model. Also, it is know~\cite{LeontAnt,leontaris} that the gauge 
symmetry $G_{PS}$ can arise from the weakly coupled heterotic string 
as well.
\vskip 0.25cm
Hybrid inflation~\cite{linde} has been extensively
studied~\cite{hybird1,hybird2, Laz1} in the case
of the SUSY model based on the gauge group $G_{LR}$. However, in trying to
extend this scheme to $G_{PS}$, we encounter the following difficulty. The
spontaneous symmetry breaking of $G_{PS}$ to the standard model gauge
group  $G_{SM}$ leads to the existence of topologically stable magnetic
monopole solutions. This is due to the fact that the second homotopy
group of the vacuum manifold $\pi_2(G_{PS}/G_{SM})$ is non-trivial and
equal to the set of integers $Z$. These monopoles carry two units of Dirac 
magnetic 
charge~\cite{magg}. Inflation is terminated abruptly when the system
reaches a critical point (instability) on the inflationary trajectory and
is followed by  a `waterfall' regime during which the spontaneous breaking
of $G_{PS}$ occurs. The appropriate Higgs fields develop their non-zero
vacuum expectation values (vevs) starting from zero and they can end up
at any point of the vacuum manifold with equal probability. As a
consequence, magnetic monopoles are copiously
produced~\cite{Panagiotakopoulos} by the Kibble mechanism~\cite{Kibble}
leading to a cosmological disaster.
\vskip 0.25cm
   In this paper, we propose a specific SUSY model based on $G_{PS}$ which
avoids this cosmological catastrophe.
This is achieved by including a non-renormalizable term in the
part of the superpotential involving the inflaton system and
causing the breaking of $G_{PS}$. It is worth mentioning that an
analogous non-renormalizable term was also used in
Ref.\cite{Panagiotakopoulos} for the same purpose. In that case,
however, the leading renormalizable term was eliminated by
imposing a discrete symmetry. Here, we keep this leading term
along with the non-renormalizable contribution. The picture that
emerges  turns out to be considerably different. In particular,
there exists a non-trivial (classically) flat direction along
which $G_{PS}$ is spontaneously broken with the appropriate Higgs
fields acquiring constant values. This direction can be used as
inflationary trajectory with the necessary inclination obtained
from one-loop radiative corrections~\cite{DvaSha} in contrast to
the model of Ref.\cite{Panagiotakopoulos}, where a classical
inclination was present. Another difference is that here the
termination of inflation is abrupt (as in the original hybrid
inflationary scenario) and not smooth as in
Ref.\cite{Panagiotakopoulos}. Nevertheless, no magnetic monopoles
are formed in this transition since $G_{PS}$ is already broken
during inflation. \vskip 0.25cm 

We show that, for a certain range
of parameters, the system always passes from the above mentioned
inflationary trajectory before falling into the SUSY vacuum. Thus,
the magnetic monopole problem is solved for all initial
conditions. It is interesting to note that the idea of breaking
the gauge symmetry before (or during) inflation in order to avoid monopoles
was also employed in Ref.\cite{masiero}. However, the monopole problem
was solved only for a certain (wide) class of initial values of the
fields. \vskip0.25cm

The constraints on the quadrupole anisotropy of the cosmic microwave
background radiation from the cosmic background explorer
(COBE)~\cite{Cobe} measurements can be easily met with natural values (of
order $10^{-3}$) of the relevant coupling constant and a grand
unification theory (GUT) scale $M_{GUT}$ close to (or somewhat smaller 
than) the SUSY GUT scale. Note that the  mass scale in the model of
Ref.\cite{DvaSha}, which uses only renormalizable couplings in the
inflationary superpotential, is considerably smaller.
   Our model possesses a number of other interesting features too. The
$\mu$-problem of MSSM is solved~\cite{PQmu} via a Peccei-Quinn
(PQ) symmetry which also solves the strong CP problem. Although
the baryon ($B$) and lepton ($L$) numbers are explicitly violated,
the proton life time is considerably higher than the present
experimental limits. Light neutrinos acquire masses by the seesaw
mechanism and the baryon asymmetry of the universe can be
generated through a primordial leptogenesis~\cite{yanagita}. The
gravitino constraint~\cite{khlopov} on the reheat temperature, the
low deuterium abundance limits~\cite{deuterium} on the baryon
asymmetry of the universe and the requirement of almost maximal
$\nu_{\mu}-\nu_{\tau}$ mixing from SuperKamiokande~\cite{japan}
can be met for  $\mu$- and $\tau$-neutrino masses restricted by
SuperKamiokande and the  large angle MSW solution of the solar
neutrino puzzle respectively. The required values of the relevant
coupling constants are more or less natural. \vskip 0.25cm
   The plan of the paper is as follows. In Sec.2, we introduce our
SUSY model which is based on the gauge group $G_{PS}$ and motivate
the inclusion of a non-renormalizable coupling in the inflaton
sector of the theory. The full superpotential and its global
symmetries are then discussed together with the solution of the
$\mu$-problem via the PQ symmetry of the model. In Sec.3, the
hybrid inflationary scenario is studied in detail in this model.
The structure of the potential is carefully analyzed. We calculate
the one-loop radiative corrections along the inflationary
trajectory by first deriving (see Appendix) the mass spectrum of the
theory
during inflation. The parameters of the model are then restricted
by employing COBE results. In Sec.4, we discuss the reheating
process following inflation, neutrino masses and mixing and
baryogenesis via leptogenesis. We show that all the relevant
constraints can be satisfied with natural values of the coupling
constants. We summarize our conclusions in Sec.5.

\section{ A SUSY ${\bf SU(4)_c \times SU(2)_L \times SU(2)_R}$ 
model}
In the SUSY PS model, the left-handed quark and lepton
superfields are accommodated in the following
representations:
\begin{eqnarray}
F_i &=& (4,2,1) \equiv \left(\begin{array}{cccc}
                       u_i & u_i & u_i & \nu_i\\
                      d_i & d_i & d_i & e_i
                      \end{array}\right) , \nonumber\\
F^c_i &=& (\bar{4},1,2) \equiv \left(\begin{array}{cccc}
                       u^c_i & u^c_i & u^c_i & \nu^c_i\\
                      d^c_i & d^c_i & d^c_i & e^c_i
                      \end{array}\right) ,
\end{eqnarray}
where the subscript $i=1,2,3$ denotes the family
index~\cite{pati}. The $G_{PS}$ gauge symmetry can be
spontaneously broken to $G_{SM}$ by a pair of Higgs superfields
\begin{eqnarray}
H^c &=& (\bar{4},1,2) \equiv \left(\begin{array}{cccc}
                       u^c_H & u^c_H & u^c_H & \nu_H^c\\
                       d^c_H & d^c_H & d^c_H & e^c_H
                      \end{array}\right) , \nonumber\\
\bar{H}^c &=& (4,1,2) \equiv \left(\begin{array}{cccc}
                       \bar{u}^c_H & \bar{u}^c_H & \bar{u}^c_H &
                       \bar{\nu}_H^c\\
                      \bar{d}^c_H & \bar{d}^c_H & \bar{d}^c_H & \bar{e}^c_H
                      \end{array}\right)
\end{eqnarray}
acquiring non-vanishing vevs in the right-handed neutrino
direction, $\vert \langle \nu_H^c \rangle \vert$ , $\vert \langle
\bar{\nu}_H^c \rangle \vert \neq 0$. \vskip 0.25cm

The two low energy Higgs doublets of the MSSM
are contained in the following representation:
\begin{eqnarray}
h = (1,2,2) \equiv \left(\begin{array}{cc}
                      h_2^+ & h_1^0 \\
                      h_2^0 & h_1^-
                      \end{array}\right).
\end{eqnarray}
After $G_{PS}$ breaking, the bidoublet Higgs field $h$ splits into
two Higgs doublets $h_1$, $h_2$, whose neutral components
subsequently develop weak vevs $\langle h^0_1 \rangle = v_1$ and
$\langle h^0_2 \rangle = v_2$ with 
$\tan \beta=v_2/v_1$. 
\vskip 0.25cm

The breaking of $G_{PS}$ can be achieved by introducing a gauge
singlet superfield $S$, which has a trilinear (renormalizable)
coupling to $H^c$, $\bar{H}^c$. The resulting scalar potential
automatically possesses an in-built (classically) flat direction
along which inflation can take place~\cite{Cop} with the system
driven by an inclination from one-loop radiative
corrections~\cite{DvaSha}. The $G_{PS}$ gauge symmetry is restored
along this trajectory and breaks spontaneously only at the end of
inflation when the system falls towards the SUSY vacua. This
transition leads to a cosmologically unacceptable copious
production of doubly charged magnetic
monopoles~\cite{Panagiotakopoulos}. One way to resolve this
problem is to use, for inflation, another flat direction in which
$G_{PS}$ is already broken. Such a trajectory naturally appears if
we include the next order non-renormalizable superpotential
coupling of $S$ to $H^c$, $\bar{H}^c$ too. We find that, together
with the usual flat direction with unbroken $G_{PS}$, an extra
flat trajectory along which $G_{PS}$ is spontaneously broken to
$G_{SM}$ emerges. The termination of inflation can then take place
with $G_{PS}$ already broken and no monopoles being produced. 
\vskip 0.25cm

   An important issue is the generation of the $\mu$-term of MSSM.
This could be easily achieved~\cite{shafi2} by coupling $S$ to the
electroweak Higgs superfields and using the fact that $S$, after
gravity-mediated SUSY breaking, develops a vev. However, this is
not totally satisfactory since the inflaton decays into electroweak 
Higgs superfields via an unsuppressed (renormalizable) coupling.
As a consequence, the gravitino constraint~\cite{khlopov} on the
reheat temperature implies~\cite{Laz2} unnaturally small
values for the relevant coupling constants (of order $10^{-6}$ or
so). We thus prefer to follow here Ref.\cite{PQmu} and impose a
PQ symmetry on the superpotential by introducing a pair of gauge
singlet superfields $N$, $\bar{N}$. The PQ breaking occurs at an
intermediate scale by the vevs of $N$, $\bar{N}$ and the
$\mu$-term is generated via a non-renormalizable coupling of $N$ 
and $h$. The inflaton can be made to decay into
right-handed neutrinos by introducing into the scheme quartic
(non-renormalizable) superpotential couplings of $\bar{H}^c$ to 
$F^c_i$ and the gravitino constraint can be satisfied with more 
natural values of the parameters. It should be noted that the 
presence of these quartic terms is anyway 
necessary for generating intermediate scale masses for the 
right-handed neutrinos and, thus, masses for the light 
neutrinos via the seesaw mechanism. Finally, in order to give
superheavy masses to $d^c_H$ and $\bar{d}^c_H$, we
introduce~\cite{LeontAnt} an $SU(4)_c$ 6-plet superfield
$G=(6,1,1)$ which, under $G_{SM}$, splits into
$g^c=(\bar{3},1,1/3)$ and $\bar{g}^c=(3,1,-1/3)$. 
\vskip 0.25cm

The superpotential of the model, which incorporates all the
above couplings, is
\begin{eqnarray}
 W & = & \kappa S (\bar{H}^c H^c - M^2)- \beta S{(\bar{H}^c H^c)^2
\over M_S^2} \nonumber \\ &+& \lambda_1 \frac{N^2 h^2}{M_S} +
\lambda_2 \frac{N^2 \bar{N}^2}{M_S} + \lambda_{ij} F^c_i F_j h
\nonumber \\ &+& \gamma_i \frac{\bar{H}^c \bar{H}^c}{M_S} F_i^c
F_i^c + a G H^c H^c + b G \bar{H}^c \bar{H}^c ,\label{eq:superpot}
\end{eqnarray}
where $M_S \sim 5 \times 10^{17}$ GeV is a superheavy string
scale. Also, $M$, $\kappa$, $\lambda_{1,2}$, $\gamma_i$, $a$ 
and $b$ can be made positive by field redefinitions, while $\beta$ 
is chosen positive for simplicity (it could be genuinely complex).
Here, we are in the basis where $\gamma$'s are diagonal.
\vskip 0.25cm

Note that the existence of the third (non-renormalizable) coupling 
in the right hand side of Eq.(\ref{eq:superpot}) is an automatic 
consequence of the first two couplings which constitute the standard 
superpotential for hybrid inflation. Indeed, the operator 
$\bar{H}^c H^c$ is neutral under all the R and non-R symmetries of 
this standard superpotential and, thus, the above coupling, which is 
crucial for our inflationary scheme, cannot be forbidden. Our only 
assumption is then that the dimensionless coefficient $\beta$ of 
this non-renormalizable coupling is of order unity so that it can be 
comparable to the trilinear coupling in the standard superpotential 
whose coefficient $\kappa$ is typically relatively small 
($\sim 10^{-3}$). This non-renormalizable coupling can then play 
an important role in the inflationary scenario. The non-renormalizable 
couplings $S(\bar{H}^cH^c)^n/M_S^{2(n-1)}$ with $n\geq 3$ are 
also allowed. They are, however, subdominant to the leading 
non-renormalizable coupling (with $n=2$) in the relevant region of 
the field space even if their coefficients are of order one.
The R symmetry (see below) of the standard superpotential, extended 
to higher orders, is needed for avoiding terms non-linear in $S$ 
which can destroy inflation. The couplings, then, permitted between 
$S$, $H^c$, $\bar{H}^c$ are the ones already mentioned modulo 
arbitrary multiplications by $(H^c)^4$, $(\bar{H}^c)^4$. The 
new couplings do not affect the potential for inflation.
\vskip 0.25cm

In addition to $G_{PS}$, the superpotential in Eq.(\ref{eq:superpot})
possesses two global anomalous symmetries, a R symmetry $U(1)_R$ and a 
PQ symmetry $U(1)_{PQ}$. The R and PQ charges of the superfields are 
assigned as follows:
\begin{eqnarray}
R :~~H^c(0), \bar{H}^c(0), S(1), G(1), F({1/2}), F^c({1/2}),
N({1/2}), \bar{N}(0), h(0); \nonumber \\ PQ :~~H^c(0),
\bar{H}^c(0), S(0), G(0), F({-1}), F^c({0}), N(-1), \bar{N}(1),
h(1).
\end{eqnarray}
To avoid undesirable mixing of $F$ and $h$ or $F^c$ and $H^c$, we
also impose a discrete $Z_2^{mp}$ symmetry (know as `matter parity'),
under which $F$ and $F^c$ change sign. \vskip 0.25cm

Additional superpotential terms allowed by the symmetries of the model are
\begin{equation}
\hspace{-.75cm}F F H^c H^c \bar{N}^2,\hspace{0.5cm} F F H^c H^c h
h,\hspace{0.5cm} F F \bar{H}^c \bar{H}^c \bar{N}^2,\hspace{0.5cm}
F F \bar{H}^c \bar{H}^c h h,  \hspace{0.5cm} F^c F^c H^c H^c , 
\label{terms}
\end{equation}
modulo arbitrary multiplications by non-negative powers of the 
combinations $H^c\bar{H}^c$, $(H^c)^4$, $(\bar{H}^c)^4$ (this 
applies to the terms in Eq.(\ref{eq:superpot}) too). Note that the 
$SU(4)_c$ indices in all couplings except the last three in 
Eq.(\ref{terms}) and the combinations $(H^c)^4$, $(\bar{H}^c)^4$ 
are contracted between $4$'s and $\bar{4}$'s, while in these terms 
and combinations the four $SU(4)_c$ indices of the $4$'s or $\bar{4}$'s 
are contracted with an $\epsilon_{ijkl}$. The soft SUSY breaking and
instanton effects explicitly break $U(1)_R$ to $Z_2$, under which
$N\rightarrow -N$, and $U(1)_{PQ}$ to $Z_6$. These two discrete
symmetries are spontaneously broken by the vevs of $N, \bar{N}$
and would create a domain wall problem if the PQ transition took
place after inflation. When $H^c$, $\bar{H}^c$, $N$ and $\bar{N}$
acquire non-vanishing vevs, the symmetry which is left unbroken is
$G_{SM} \times Z_2^{mp}$. \vskip 0.25cm

We can assign baryon number $1/3(-1/3)$ to all color triplets
(antitriplets). Recall that there are (anti)triplets not only in $F,
F^c$ but also in $H^c, \bar{H}^c, G$. Lepton number is then
defined via $B-L$. $B$ (and $L$) violation comes from
the last three terms in Eq.(\ref{terms}) (and the combinations 
$(H^c)^4$, $(\bar{H}^c)^4$) which give couplings like \
$u^c\ d^c\ d^c_H\ \nu^c_H \ (\mathrm{or}\ u^c\ d^c\ u^c_H\
e^c_H)$,\ $u\ d\ \bar{d}^c_H\ \bar{\nu}^c_H 
\ (\mathrm{or}\ u\ d\ \bar{u}^c_H\ \bar{e}^c_H)$ with 
appropriate coefficients. Also,
the terms $G H^c H^c$ and $G \bar{H}^c \bar{H}^c$ give 
rise to the $B$ (and $L$) violating couplings
$g^c\ u^c_H\ d^c_H$,\ $\bar{g}^c\
\bar{u}^c_H\ \bar{d}^c_H$. All other combinations are $B$ (and
$L$) conserving since $4$'s are contracted with $\bar{4}$'s.
\vskip 0.25cm

The dominant contribution to proton decay comes from effective 
dimension five operators generated by one-loop diagrams with two of 
the $u^c_H$, $d^c_H$ or one of the $u^c_H$, $d^c_H$ and one of the 
$\nu^c_H$, $e^c_H$ circulating in the loop. The amplitudes 
corresponding to these operators are estimated to be at most of order 
$m_{3/2}M_{GUT}/M_S^3 \stackrel{_{<}}{_{\sim }} 10^{-34}~ 
{\rm{GeV}}^{-1}$ ($m_{3/2}$ 
is the gravitino mass). This makes the proton practically stable.
Furthermore, the dominant contribution to the Majorana mass term
of light neutrinos comes from $FFH^cH^chh$ and is utterly small.
So the seesaw mechanism is the only source of light neutrino masses.
\vskip 0.25cm

The $\mu$-term is generated, as mentioned, by a non-renormalizable
superpotential coupling which contains the electroweak Higgs and
the $N$ superfield after the breaking of $U(1)_{PQ}$ by $\langle
N\rangle$, $\langle \bar{N}\rangle$.  The relevant part of the
scalar potential for the PQ breaking is given by~\cite{PQmu}
\begin{equation}
V_{PQ} = 2 \vert N \vert^2 m_{3/2}^2 \left( 4 \lambda_2^2 \frac{\vert N
\vert^4 }{
m_{3/2}^2 M_S^2} - \vert A \vert \lambda_2 \frac{\vert N \vert^2
}{m_{3/2} M_S} + 1 \right),
\label{potn0}
\end{equation}
where $A$ is the dimensionless coefficient of the soft SUSY breaking term 
corresponding to the $N^2\bar{N}^2$ term in Eq.(\ref{eq:superpot}). 
Here, the phases
$\epsilon$, $\theta$ and $\bar{\theta}$ of $A$, $N$ and $\bar{N}$
are taken to satisfy the relation $\epsilon + 2 \theta + 2
\bar{\theta} = \pi$ and $\vert N \vert$, $\vert \bar{N} \vert$ are
assumed equal which minimizes the potential. For $\vert A \vert >
4$, the absolute minimum of this potential is given by~\cite{PQmu}
\begin{equation}
\vert \langle N \rangle \vert = \vert \langle \bar{N} \rangle \vert =
(m_{3/2} M_S)^{1/2} \left(\frac{\vert A \vert + \sqrt{\vert A \vert^2
-12}}{12\lambda_2}\right)^{1/2}.
\end{equation}
Hence the PQ symmetry breaking scale is of order $\sqrt{m_{3/2}
M_S} \simeq 10^{10} - 10^{11}$ GeV and the $\mu$-term of the MSSM
is $\sim m_{3/2}$ as desired. 
\vskip 0.25cm

Note that the zero temperature PQ potential (in Eq.(\ref{potn0})),
shown in Fig.1, possesses two local minima,  the trivial one at
$\vert N \vert =0$ and the PQ minimum which, for $\vert A\vert
>4$, is the absolute minimum . These minima are separated by a
sizable potential barrier which prevents a successful transition
from the trivial to the PQ vacuum. Taking the one-loop temperature
corrections~\cite{jackiw} to the potential into account, one can
show that the PQ vacuum remains the absolute minimum at least for
temperatures below the reheat temperature $T_r \sim 10^9$ GeV.
The trivial vacuum is still protected by a potential barrier. We,
thus, conclude that if, after inflation, the system emerges in the
trivial vacuum the completion of the PQ transition will be
practically impossible. We are obliged to assume that the PQ
symmetry is already broken before or during inflation. The PQ
vacuum then remains stable after inflation and reheating. There is
yet another reason which disfavors a PQ transition after
inflation. The vevs of $N$, $\bar N$ break spontaneously the $Z_2$
symmetry ($N\rightarrow -N$) and the $Z_6$ subgroup of $U(1)_{PQ}$
which is left unbroken by instantons. This would lead to
disastrous domain walls in the universe. \vskip 0.25cm

In concluding this section, we emphasize that the extra couplings 
which we incorporated into our basic inflationary superpotential (see 
Eq.(\ref{eq:superpot})) are simple, quite general, and well-motivated. 
In particular, the couplings $N^2 \bar{N}^2$, $N^2 h^2$ were included 
in order to solve the $\mu$-problem. However, they also provide a 
solution to the strong CP problem. Moreover, the presence of 
$\bar{H}^c\bar{H}^cF_i^cF_i^c$ is necessary for the generation of 
the right-handed neutrino masses. It is an extra benefit, though, that, 
via these same couplings, the inflaton decays to right-handed 
neutrinos, thereby leading to a successful and `natural' 
(no tiny couplings) reheating of the universe with the observed 
baryon asymmetry generated via leptogenesis (see below). Finally, 
$G H^c H^c$, $G\bar{H}^c\bar{H}^c$ were added for merely giving 
superheavy masses to $d^c_H$, $\bar{d}^c_H$. It is worth mentioning 
that the emerging theory is `natural' in the sense that all couplings 
which are consistent with a simple and general set of global symmetries 
($U(1)_R$, $U(1)_{PQ}$ and $Z_2^{mp}$) can be allowed without 
jeopardizing our inflationary scenario or leading into trouble with 
other requirements such as proton stability.    
\vskip 0.25cm

\section{ The inflationary scenario}
The part of the superpotential in Eq.(\ref{eq:superpot})
which is relevant for inflation is
given by
\begin{equation}
\delta W = \kappa S (\bar{H}^c H^c - M^2)- \beta \frac{S
(\bar{H}^c H^c)^2}{M_S^2}, \label{eq:susyinfl}
\end{equation}
where $M$ is a superheavy mass scale close to the GUT scale.
Note that the rest of the superpotential in  Eq.(\ref{eq:superpot}) does
not affect the inflationary dynamics. The scalar potential obtained from
$\delta W$ is given by
\begin{eqnarray}
V &=& \left \vert \kappa (\bar{H}^c H^c  - M^2) -\beta
\frac{(\bar{H}^c
H^c)^2}{M_S^2} \right \vert^2 + \kappa^2 \vert S \vert^2 (|H^c|^2
+|\bar{H}^c|^2)
\left \vert 1 - \frac{2
\beta}{\kappa M_S^2} \bar{H}^cH^c \right \vert^2 \nonumber\\
        &+&  {\rm{D}}-\mathrm{terms}, \label{eq:infpot}
\end{eqnarray}
where the scalar components of the superfields are denoted by the
same symbols as the corresponding superfields. Vanishing of the
D-terms is achieved with $\vert \bar{H}^c \vert = \vert H^c
\vert $ ($\bar{H}^c$ ($H^c$) lies in the $\bar{\nu}^c_H$
($\nu^c_H$) direction). Restricting ourselves to this direction 
and performing an appropriate R transformation, we can bring the complex 
field $S$ to the real axis, $S = \sigma/ \sqrt{2}$, where $\sigma$ 
is a normalized real scalar field. An $\vert S
\vert$-independent flat trajectory suitable for hybrid inflation
driven by radiative corrections is obtained in the direction
$\mathrm{arg}(H^c)+\mathrm{arg}(\bar{H}^c)=0$ along which the
potential takes the form
\begin{equation}
V = \left[ \kappa (\vert H^c \vert^2 - M^2) -\beta \frac{\vert H^c
\vert^4}
{M_S^2} \right]^2 + \kappa^2
\sigma^2 \vert H^c \vert^2 \left[1 - \frac{2 \beta}{\kappa M_S^2}
\vert H^c \vert^2 \right]^2
.\label{eq:inflpot}
\end{equation}
We can now rewrite this potential in terms of the dimensionless
variables
$$
w= \frac{\vert S \vert }{M}, \hspace{1.5cm} y= \frac{\vert H^c \vert}{M}.
$$
We get
\begin{equation}
\tilde{V} = \frac{V}{\kappa^2 M^4} = (y^2 - 1 - \xi y^4)^2 + 2 w^2
y^2 (1 - 2 \xi y^2)^2 , \label{vtilde}
\end{equation}
where $\xi= \beta M^2 / \kappa M_S^2$. This potential is a simple 
extension of the standard potential for SUSY hybrid 
inflation (obtained from Eq.(\ref{vtilde}) by putting $\xi=0$) 
and appears generically in a wide class of models which incorporate the 
leading non-renormalizable correction to the standard hybrid inflationary 
superpotential. (Recall that this correction is naturally present 
as we explained.) It is, thus, interesting to study in some detail the 
structure of this potential and the new inflationary possibilities which 
typically emerge.   
\vskip 0.25cm

For constant $w^2$, the potential in Eq.(\ref{vtilde})
has the following extrema
\begin{eqnarray}
y_1 &=& 0, \\ y_2 &=& \sqrt{\frac{1}{2 \xi}},\\ y_{3\pm}^2 &=&
\frac{1}{2 \xi}\left[(1 - 6 \xi w^2)\pm \sqrt{(1 - 6 \xi w^2)^2 -
4\xi(1 - w^2)} \right]. \label{y3}
\end{eqnarray}
Note that the first two extrema are $\sigma$-independent. As it
turns out, $y_1$ is a local minimum (maximum) if $w> (<) 1$, while
$y_2$ is a local minimum (maximum) if $w^2>(<)$ $\rho_0 = 1/8\xi-1/2$.
Inflation will take place when the system is trapped along the
$y_2$ minimum.  We restrict ourselves to $\xi < 1/4$ since, in this case,
the inflationary trajectory (at $y_2$) is destabilized in the direction of
the real part of $H^c \bar{H}^c$ before $w$ reaches zero. Inflation can
then be terminated with the system falling towards the SUSY vacua
(see below) following the direction $\mathrm{arg}(H^c) +
\mathrm{arg}(\bar{H}^c)=0$, where the potential is given by
Eq.(\ref{vtilde}). The potential at $y_2$ is $\tilde{V}_2 = (1/4\xi-1)^2$,
while at $y_1$ is $\tilde{V}_1 =1$. So that, for $\xi >(<) 1/8$, the
extremum at $y_1$ lies higher (lower) than the one at $y_2$.
\vskip 0.25cm
For $1/4> \xi>1/7.2$, the discriminant $\Delta$ appearing
under the  square root in Eq.(\ref{y3}) is negative if $w^2$ lies between
the
positive  numbers $\rho_{\pm}=(2 \pm \sqrt{36 \xi - 5})/18\xi$ and
non-negative otherwise. So, for $\rho_{-} < w^2 <\rho_{+}$, the extrema at
$y_{3\pm}$ do not exist.  Note, however, that $\Delta \geq 0$ does not
necessarily guarantee the presence of these extrema. This requires that,
in addition, the right hand side of Eq.(\ref{y3}) is non-negative. An
important
ingredient on which this requirement depends is the sign of the
expression $1-6\xi w^2$ which is positive (negative) for $w^2 <(>)
\rho_1=1/6\xi$. One can show that, for $1/4 > \xi >1/7.2$,
$0< \rho_0 <\rho_{-} <\rho_{+} \leq 1$ (equality holds
for $\xi=1/6$) and $\rho_0 < \rho_1$. Also, $\rho_{+} >(<) \rho_1$ for
$\xi>(<)1/6$.
\vskip 0.25cm

For $\xi \leq 1/7.2$, we always have $\Delta \geq 0$. In addition, 
note that
$\rho_0 <(>) 1$ for $\xi >(<) 1/12 $, $\rho_1 <(>) 1$ for $\xi >(<) 1/6$
and $ 0< \rho_0 < \rho_1$ always. It is interesting to point out that, at
$w^2=\rho_0$, above which the extremum at $y_2$ turns into a local
minimum, $y_{3+}$ coincides with $y_2$. The minima at $y_{3\pm}$, for
$w^2=0$, become supersymmetric, i.e., $\tilde{V}(w^2=0,
y=y_{3\pm})=0$. For reasons to become obvious later,
we consider the $y_{3-}$ minimum at $w^2=0$ as the relevant
SUSY vacuum of the theory.
\vskip 0.25cm

    Taking into account these facts, we can distinguish five cases
with qualitatively different structure of the potential:

   $(i)$ For $1/4> \xi > 1/6$, we have $0 <
\rho_0< \rho_{-}< \rho_1<\rho_{+}<1$.
One can then show that, for fixed $w^2>1$, there exist two local
minima at $y_1$ and $y_2$ (the interesting inflationary trajectory)
and a local maximum at $y_{3+}$ between them (see Fig.2). For $w^2$
between $1$ and $\rho_{-}$, the trivial extremum at $y_1$ becomes a local
maximum and the extremum at $y_{3+}$ disappears (see Fig.3). In this range
of $w^2$, the system can freely fall into the desirable (inflationary)
minimum at $y_2$
even if it was initially along the trivial trajectory at $y_1$
(remember that the extremum at $y_2$ lies lower than the one at $y_1$
in this case). As we further decrease $w^2$ to become smaller than
$\rho_{-}$, a pair of two new extrema, a local minimum at $y_{3-}$ and
a local maximum at $y_{3+}$, are created between $y_1$ and $y_2$. As $w^2$
crosses $\rho_0$, the local maximum at $y_{3+}$ crosses $y_2$ becoming a
local minimum (see Fig.4). At the same time, the local minimum at $y_2$
turns  into a local maximum and inflation is terminated with the system
falling into the local minimum at $y_{3-}$ which at $w^2=0$ develops
into a SUSY vacuum (see below).
\vskip 0.2cm

$(ii)$ For $1/6> \xi >1/7.2$, we have
$0 < \rho_0 < \rho_{-} < \rho_{+} < 1 <\rho_1$.
The situation for $w^2>1$ and $w^2<\rho_{-}$ is similar to the previous
case. For $w^2$ between $1$ and $\rho_{+}$, the $y_1$ extremum becomes a
local
maximum and a local minimum at $y_{3-}$ appears between $y_1$ and
$y_{3+}$.
As $w^2$ decreases below $\rho_{+}$, the extrema at $y_{3\pm}$ disappear
and there exists no obstacle for the system to fall to $y_2$ even if it
was
initially at $y_1$. The extrema at $y_{3\pm}$ reappear as $w^2$ becomes
smaller than $\rho_{-}$.
\vskip 0.2cm

     $(iii)$ For $1/7.2 > \xi > 1/8$, $0< \rho_0 < 1< \rho_1$. The
behavior of the
potential for $w^2 > 1$ and $ w^2 < \rho_0$  is similar to the previous
cases. For $1 > w^2 > \rho_0$, however, the extremum at $y_1$ becomes a
local  maximum and a local minimum at $y_{3-}$ appears between $y_1$ and
$y_{3+}$. Notice that, in this case, although the extremum at $y_2$ lies
lower than the one at $y_1$, there is no range of $w^2$ where the system
can fall into $y_2$ if it was initially at $y_1$. Instead, it ends up
directly in $y_{3-}$ from $y_1$ and monopoles can be copiously produced.
Of course, if the system happens to be at $y_2$ from the beginning,
there is no production of monopoles.
\vskip 0.2cm

      $(iv)$ For $1/8 > \xi > 1/12$, the situation is exactly as in case
$(iii)$ with the only difference that the extremum at $y_2$ now lies
higher than  the one at $y_1$.
\vskip 0.2cm

       $(v)$ For $1/12 > \xi$, we have $0 < 1 < \rho_0<\rho_1$. It turns
out that, for $w^2 > \rho_0$, the local minima at $y_1$ and $y_2$ (which
lies higher) are again  separated by a local maximum at $y_{3+}$. As $w^2$
crosses $\rho_0$, the $y_{3+}$ local maximum turns into minimum and
crosses $y_2$ which becomes a local
maximum. There is then no obstacle to keep the system from falling into
$y_1$ even if it was at $y_2$. Subsequently, when $w^2$ becomes smaller
than $1$, $y_1$ turns into a local maximum and the system falls into
$y_{3-}$ with a  copious production of magnetic monopoles.
\vskip 0.25cm

We will restrict ourselves here to the first two cases above $(1/4
> \xi > 1/7.2)$. We saw that, in these cases, even if the system
starts along the trivial valley at $y_1$, it always falls into
the (classically) flat direction at $y_2$. The relevant part of
inflation can then take place along this trajectory with the
inflaton being driven by radiative corrections~\cite{DvaSha}. So,
$G_{PS}$ is already broken during inflation and
there is no production of magnetic monopoles at the end of
inflation where the system falls into the $y_{3-}$ minimum. Case
$(iii)$ could also solve the monopole problem provided the system
starts at $y_2$. Case $(iv)$, although quite similar to case
$(iii)$, is more tricky requiring further study. The reason is
that, since $y_2$ lies higher than $y_1$, the system oscillates
over the local maximum at $y_1$ after falling from $y_2$. Finally,
case $(v)$ is always unacceptable since the system, for all
initial conditions, falls to $y_{3-}$ from $y_1$ and the copious
production of monopoles is unavoidable. \vskip 0.25cm

As we already mentioned, after inflation ends, the system falls 
into the minimum at $y_{3-}$ which, at $w^2=0$, develops into the final 
SUSY vacuum of the theory. However, the system could, in 
principle, fall into the minimum at $y_{3+}$ which appears only after 
the instability of the inflationary trajectory at $w^2=\rho_{0}$ is 
reached. (The minimum at $y_{3+}$ also develops into a SUSY 
vacuum at $w^2=0$.) We will argue that this does not happen. For 
the values of the parameters used here, the potential barrier 
separating the inflationary path at $y_2$ and the minimum at $y_{3-}$ 
is considerably reduced in the last e-folding or so. (The 
peak of this barrier coincides with the maximum at $y_{3+}$.) As a 
consequence, the rate per unit volume and time of forming bubbles 
of the $y_{3-}$ minimum ceases to be exponentially suppressed. An 
order of magnitude estimate then shows that the decay of the false 
vacuum at $y_2$ to the minimum at $y_{3-}$ is completed within a 
fraction of one e-folding. This happens before the appearance of the 
minimum at $y_{3+}$, i.e., before the system reaches the critical 
point at $w^2=\rho_{0}$ (but very close to this point). 
Moreover, in the 
last stages of inflation, the above barrier is small enough to be
overcome by the inflationary density perturbations. This can also 
accelerate the completion of this transition.
\vskip 0.25cm

To avoid confusion we should mention here that $\xi$ is not an
extra free parameter. It depends on the coupling $\kappa$ and the
superheavy mass scale $M$ (we put $\beta=1$). The values of
$\kappa$ and $M$  will be related by calculating the quadrupole
anisotropy of the cosmic microwave background radiation $(\delta T/T)_Q$
as a function of the number of e-foldings of our present horizon
$N_Q$ and compare it with the measurements of COBE~\cite{Cobe}.
The parameter $\xi$ then becomes a function of the basic coupling
constant of the scheme $\kappa$. So, searching for solutions with
$\xi$ in the desirable range and also satisfying all the other
requirements which we will discuss below is a highly non-trivial task.
\vskip 0.25cm

         As already mentioned, the interesting part of inflation
takes place when the system is trapped along the trajectory at
$y_2$. Inflation is driven by the constant classical energy
density on this trajectory which also breaks SUSY. This
breaking gives rise to non-trivial radiative
corrections~\cite{DvaSha,RJ} which lift the (classical) flatness
of this trajectory producing a necessary inclination for driving
the inflaton towards the SUSY vacua. As will be seen later, the 
slow-roll conditions~\cite{LythLid} are satisfied and inflation 
continues essentially till $w^2$ reaches $\rho_0$,
where the inflationary trajectory is destabilized. To calculate
the one-loop radiative corrections at $y_2$ we need to construct
the mass spectrum of the theory on this path where both $G_{PS}$
and SUSY are broken. Details of the calculation can be
found in the Appendix. We summarize the results in Table 1.

\begin{table} [ht]
\begin{center}
\begin{tabular}{c c}
\hline \hline Fields  & Squared Masses \\ \hline 2
real scalars  & $4 \kappa^2 \vert S \vert^2 \mp 2
\kappa^2 M^2 ({1\over 4 \xi}-1)$ \\ \hline
1 Majorana fermion & $4 \kappa^2 \vert S \vert^2$\\\hline
1 real scalar & $5 g^2 v^2/2$ \\ \hline
1 gauge boson  & $5 g^2 v^2/2$ \\ \hline
1 Dirac fermion & $5 g^2 v^2/2$\\\hline
8 real scalars & $g^2 v^2$ \\\hline
8 gauge bosons & $g^2 v^2$ \\ \hline
8 Dirac fermions & $g^2 v^2$\\ \hline
6 complex scalars & $4 a^2 v^2$  \\ \hline
3 Dirac fermions & $4 a^2 v^2$\\ \hline
6  complex scalars & $4 b^2 v^2$  \\ \hline
3 Dirac fermions & $4 b^2 v^2$\\\hline \hline
\end{tabular}\end{center}
{Table 1: The mass spectrum of the model as the system moves along the
inflationary trajectory at $y_2$. The parameter $v
=(\kappa M_S^2/2\beta)^{1/2}$ is the vev of $\nu^c_H$, $\bar{\nu}^c_H$ on
this trajectory and $g$ is the $G_{PS}$ gauge coupling constant.}
\end{table}
\vskip 0.3cm

Using this spectrum, we can now calculate the one-loop radiative
correction to the potential along the inflationary trajectory from
the Coleman-Weinberg formula~\cite{CW}
\begin{equation}
\Delta V = \frac{1}{64 \pi^2} \sum_i(-)^{F_i}\ M_i^4 \ln
\frac{M_i^2}{\Lambda^2}, \label{deltav}
\end{equation}
where the sum extends over all helicity states $i$, $F_i$ and
$M_i^2$ are the fermion number and squared mass of the $i$th state
and $\Lambda$ is a renormalization mass scale. We find that the
inflationary effective potential is given by
\begin{eqnarray}
V_{infl}^{eff}=\kappa^2 m^4\!\left(\!1+\!{\kappa^2\over
16\pi^2}\!\left[2 \ln {2\kappa^2\sigma^2 \over
\Lambda^2}\!+\!(z+1)^{2}
\ln(1+z^{-1})\!+\!(z-1)^{2}\ln(1-z^{-1})\right]\right),
\label{vinf}
\end{eqnarray}
where $m^2=M^2(1/4\xi-1)$ and $z=\sigma^2/m^2$. We see that the
only non-zero contributions to the effective potential come from
the $\vert S \vert$-dependent part of the spectrum (in the first 
two lines of Table 1). This is a consequence of the fact that,
along the inflationary trajectory, SUSY breaking, due to the 
presence of
non-zero vacuum energy density, occurs only in the inflaton
sector. In particular, there is mass splitting only in the
supermultiplet which contains the complex scalar field
$\theta=(\delta \nu^c_H + \delta \bar{\nu}^c_H)/\sqrt{2}$ (see
Appendix). \vskip 0.25cm

   Note that radiative corrections lift the (classical) flatness
of the inflationary trajectory providing the necessary inclination
for driving the inflaton field $S$ towards zero. It is important to
observe that although the effective potential in Eq.(\ref{vinf})
does depend on the unknown scale $\Lambda$, its inclination
(derivative with respect to $\sigma$) is $\Lambda$-independent.
This is due to the fact that the supertrace of $M^4$ ($M^2$ being
the mass squared matrix) appearing in Eq.(\ref{deltav}) is, as one
can readily deduce using the spectrum in Table 1,
$\sigma$-independent. This is an important property since
otherwise $(\delta T/T)_Q$ and $N_Q$ would depend on the unknown
mass parameter $\Lambda$. \vskip 0.25cm

   Inflation is terminated only very close to the critical
point $\sigma=m$ ($z=1$ or $w^2=\rho_0$) after which the
inflationary path is destabilized and the system falls into
$y_{3-}$. This can be checked, for all relevant values of
the coupling constants, by employing the slow-roll parameters 
$\epsilon$ and
$\eta$~\cite{LythLid}. It turns out that $\epsilon \ll 1$ for $z
\geq 1$, while $\vert \eta \vert$ exceeds unity only for $z$'s
extremely close to 1. \vskip 0.25cm

   The quadrupole anisotropy of the cosmic microwave background
radiation can be calculated to be \cite{Laz1}:
\begin{equation}
\left ( {\delta T \over T} \right )_Q \simeq \pi \left ( {32 N_Q
\over 45} \right )^{1\over 2} \left ( {m \over M_{P}} \right )^2
x_Q^{-1} y_Q^{-1} \Lambda(x_Q^2)^{-1} , \label{eq:dTT}
\end{equation}
where $M_P=1.22 \times 10^{19}$ GeV is the Planck scale,
\begin{equation}
\Lambda (z) =(z+1) \ln(1+z^{-1}) +(z-1) \ln(1-z^{-1}) ,
\end{equation}
and
\begin{equation}
y_Q^2 = \int_1^{x_Q^2} {dz\over z} \Lambda^{-1} (z), y_Q \geq 0 ,
\end{equation}
with $x_Q = \vert \sigma_Q \vert/m$, $\sigma_Q$ being the value of
$\sigma$ when our present horizon scale crossed outside the
inflationary horizon. The coupling constant $\kappa$ can be
evaluated from~\cite{Laz1}
\begin{equation}
\kappa = {4 \pi^{3\over 2} \over \sqrt{N_Q}} {m\over M_{P}} y_Q.
\label{eq:kappa}
\end{equation}
Now, using the COBE constraint, $(\delta T/T)_Q=6.6 \times
10^{-6}$~\cite{Cobe}, taking $N_Q=55$ and eliminating $x_Q$
between Eqs.(\ref{eq:dTT}) and (\ref{eq:kappa}), we can obtain $m$
and, consequently, $\xi$ and $M$ as functions of $\kappa$ (we put
$\beta=1$ and $M_S=5\times 10^{17}~{\rm{GeV}}$). The 
inflationary scale $v_{\mathrm{infl}} =
\kappa^{1/2} m$ and the spectral index of density perturbations
$n=1-6\epsilon+2\eta$ are then also found as functions of $\kappa$
and are depicted in Figs.5 and 6 respectively. \vskip 0.25cm
   The SUSY minimum can be obtained from $y_{3-}$ in
Eq.(\ref{y3})
by putting $w=0$. The common vev $v_0=\vert \langle \nu^c_H
\rangle \vert = \vert \langle \bar{\nu}^c_H \rangle \vert$ of
$H^c$ and $\bar{H}^c$ at this minimum is then given by
\begin{equation}
(\frac{v_0}{M})^2=\frac{1}{2\xi} (1-\sqrt{1-4\xi}),
\end{equation}
and is shown in Fig.7 as a function of $\kappa$.

\section{Neutrino masses and lepton asymmetry}
A complete inflationary scenario should be followed by a successful 
reheating which can generate the observed baryon asymmetry of the 
universe. We will now turn to the discussion of this reheating 
process in our model. The inflaton consists of the two complex 
scalar fields $S$ and $\theta$, which have equal masses given by
\begin{equation}
m_{\mathrm{infl}}^2 = 2 \kappa^2 v_0^2 (1-\frac{2\xi
v_0^2}{M^2})^2 .
\end{equation}
At the end of inflation, the two fields $S$ and $\theta$ oscillate
about the SUSY minimum and decay into a pair of
right-handed sneutrinos ($\nu^c_i$) and neutrinos
($\psi_{\nu^c_i}$) respectively. The masses of these (s)neutrinos
are generated, after the breaking of $G_{PS}$, by the
superpotential coupling $\gamma_i \bar{H}^c \bar{H}^c F^c_i F^c_i/M_S$
in Eq.(\ref{eq:superpot}) and turn out to be
\begin{equation}
M_i = 2 \gamma_i \frac{v_0^2}{M_S} . \label{m0}
\end{equation}
This same coupling together with the terms in
Eq.(\ref{eq:susyinfl}) constitute the part of the superpotential
which is relevant for the decay of the inflaton. We obtain the
Lagrangian terms
\begin{equation}
 L^S_{decay} = - \sqrt{2} \gamma_i \frac{v_0}{M_S}\ S^*\
\nu^c_i\ \nu^c_i\ m_{\mathrm{infl}}\ + h.c.  \label{eq:LS}
\end{equation}
for the decay of $S$ and
\begin{equation}
L^\theta_{decay} = - \sqrt{2} \gamma_i \frac{v_0}{M_S}\ \theta\
\psi_{\nu_i^c}\ \psi_{\nu_i^c}\ + h.c. \label{eq:Ltheta}
\end{equation}
for the decay of $\theta$ respectively. From Eqs.(\ref{eq:LS}) and
(\ref{eq:Ltheta}), we deduce that $S$ and $\theta$ have equal
decay widths given by
\begin{equation}
\Gamma_{S \rightarrow\ \nu^c_i \nu^c_i} = \Gamma_{\theta
\rightarrow\ \bar\psi_{\nu^c_i} \bar\psi_{\nu^c_i}}\equiv\Gamma= 
\frac{1}{8\pi}\left(\frac{M_i}{v_0}\right)^2 m_{\mathrm{infl}} ,
\end{equation}
provided that $M_i < m_{\mathrm{infl}}/2$. To minimize the number of small
couplings we assume that
\begin{equation}
M_2 < m_{\mathrm{infl}}/2 \leq M_3, \label{eq:ineq}
\end{equation}
so that the coupling $\gamma_3$ can be of order one. The inflaton then 
decays into the second heaviest right-handed neutrino superfield with mass 
$M_2$. (There always exist $\gamma_3$'s smaller than one which satisfy 
the second inequality in this equation for all relevant values of the other
parameters.) Thus the reheat temperature $T_r$ after inflation, for the
MSSM spectrum, is given by~\cite{hybird1}
\begin{equation}
T_r = \frac{1}{7} (\Gamma M_P)^{1/2} = \frac{1}{7}
\left(\frac{M_P\ m_{\mathrm{infl}}}{8\pi}\right)^{1/2}
\frac{M_2}{v_0}.
\end{equation}
The gravitino constraint~\cite{khlopov} gives an upper bound on
$T_r$ of about $10^9$ GeV for gravity-mediated SUSY
breaking with universal boundary conditions. To maximize the
naturalness of the model, we take the maximal value of $M_2$ (and
thus $\gamma_2$) allowed by the gravitino constraint. Note that
these values of $M_2$ turn out to be about two orders of magnitude
lower than the corresponding values of $m_{\mathrm{infl}}/2$ and,
thus, the first inequality in Eq.(\ref{eq:ineq}) is well
satisfied. \vskip 0.25cm

Analysis~\cite{giunti} of
the CHOOZ experiment~\cite{chooz} shows that the solar and atmospheric 
neutrino oscillations decouple, allowing us to concentrate 
on the two heaviest families. The light neutrino mass 
matrix is then given by
\begin{equation}
m_{\nu} = - \tilde{M}^{D} \frac{1}{M^R} M^{D}, \label{eq:neut1}
\end{equation}
where $M^D$ is the Dirac neutrino mass matrix with positive
eigenvalues $m^D_{2,3}$ ($m^D_2 \leq m^D_3$), and $M^R$ the Majorana
mass matrix of the right-handed neutrinos with positive
eigenvalues $M_{2,3}$ ($M_2 \leq M_3$) given in Eq.(\ref{m0}). The
two positive eigenvalues of $m_{\nu}$ will be denoted by $m_2$ (or
$m_{\nu_{\mu}}$) and $m_3$ (or $m_{\nu_{\tau}}$), with $m_2 \leq
m_3$. The determinant and trace invariance of $m_{\nu}^{\dag} m_{\nu}$ 
provide us with two constraints~\cite{Laz3} on the mass parameters
$m_{2,3}$, $m^D_{2,3}$, $M_{2,3}$ and the angle $\theta$ and phase
$\delta$ of the rotation matrix which diagonalizes the
right-handed neutrino mass matrix $M^R$ in the basis where $M^D$
is diagonal. \vskip 0.25cm

   The bounds on $m_{\nu_{\mu}}$ from the small or large angle MSW
solution of the solar neutrino puzzle are respectively $2 \times 10^{-3}\
\mathrm{eV}\ \leq m_{\nu_{\mu}} \leq 3.2 \times 10^{-3}$ eV or $3.6 \times
10^{-3} \mathrm{eV}\ \leq m_{\nu_{\mu}} \leq 1.3 \times 10^{-2}$
eV~\cite{bahcall}. As we will see below, the latter solution is favored
in our model. The $\tau$-neutrino mass is
restricted in the range $3 \times 10^{-2}$ eV $ \leq
m_{\nu_{\tau}} \leq 11 \times 10^{-2}$ eV from the results of
SuperKamiokande~\cite{japan} which also imply almost maximal
$\nu_{\mu}-\nu_{\tau}$ mixing, i.e., $\sin^{2} 2\theta_{\mu\tau}
>0.8$. Assuming that the Dirac mixing angle $\theta^D$ (i.e., the
mixing angle in the absence of right-handed neutrino 
Majorana masses) is negligible, we
find~\cite{Laz3} $\theta_{\mu\tau} \simeq \varphi$, where $\varphi$
is the rotation angle which diagonalizes $m_{\nu}$. \vskip 0.25cm

   An important constraint comes from the baryon asymmetry
of the universe. In this model, a primordial lepton asymmetry is
generated~\cite{yanagita} by the decay of the superfield $\nu^c_2$ which
emerges as the decay product of the inflaton. (This lepton
asymmetry is subsequently partially converted into baryon asymmetry by
electroweak sphalerons.) The superfield $\nu^c_2$ decays
into electroweak Higgs and (anti)lepton superfields. The resulting
lepton asymmetry is~\cite{Laz3}
\begin{equation}
\frac{n_L}{s} \simeq 1.33\  \frac{9 T_r }{16 \pi
m_{\mathrm{infl}}} \frac{M_2}{M_3}\  \frac{c^2 s^2 \sin 2 \delta
(m_3^{D\hspace{0.05cm}^2} - m_2^{D\hspace{0.05cm} ^2})^2}{v_2^2
(m_3^{D\hspace{0.05cm}^2} s^2 + m_2^{D\hspace{0.05cm} ^2} c^2)} ,
\label{eq:nls}
\end{equation}
where $s=\sin\theta$ and $c=\cos\theta$. This is related to the
baryon asymmetry $n_B/s$ by $n_L/s = - (79/28)(n_B/s)$ for the spectrum of
the MSSM~\cite{ibanez}. Thus, the low deuterium
abundance constraint~\cite{deuterium} on the baryon asymmetry of
the universe $0.017 \leq \Omega_B h^2 \leq 0.021$ gives $1.8
\times 10^{-10} \leq -n_L/s \leq 2.3 \times 10^{-10}$. \vskip
0.25cm
   Due to the presence of $SU(4)_c$ in $G_{PS}$, the Dirac mass parameter
$m^D_3$ coincides with the asymptotic value of the top quark mass.
Taking renormalization effects into account, in the context of the
MSSM with large $\tan\beta$, we
find~\cite{Laz3} $m^D_3=110-120~{\rm{GeV}}$.
\vskip 0.25cm

For each value of $\kappa$, the Majorana masses $M_{2,3}$ are fixed.
Taking
$m_{2,3}$ and $m^D_3$ also fixed in their allowed ranges, we are left with
only
three undetermined parameters $\delta$, $\theta$ and $m^D_2$ which
are further restricted by four constraints: almost maximal
$\nu_{\mu}-\nu_{\tau}$ mixing ($\sin^{2} 2 \theta_{\mu\tau}>0.8$),
the leptogenesis restriction ( $1.8 \times 10^{-10} \leq -n_L/s
\leq  2.3 \times 10^{-10}$) and the constraints from the trace and
determinant invariance of $m_{\nu}^{\dag} m_{\nu}$. It is highly non-trivial
that solutions satisfying all the above requirements can be found
with natural values of $\kappa$ (of order $10^{-3}$) and $m^D_2$ of
order $1$ GeV. Typical solutions can be constructed, for instance, for
$\kappa = 4 \times 10^{-3}$, which corresponds to $\xi \simeq
0.2$, $v_0 \simeq 1.7 \times 10^{16}$ 
GeV, $m_{\mathrm{infl}} \simeq 4.1 \times 10^{13}$ GeV, $M_2 \simeq 5.9
\times 10^{10}$ GeV and $M_3 \simeq 1.1 \times 10^{15}$ GeV 
($\gamma_3=0.5$). (Remember $\beta=1$, $M_S=5 \times 10^{17}$ GeV and 
$T_r=10^{9}$ GeV.) Taking, for example, $m_{\nu_{\mu}}= 7.6 \times
10^{-3}$ eV, $m_{\nu_{\tau}}=8 \times 10^{-2}$ eV and $m^D_3=120$ GeV,
we find $m^D_2 \simeq 1.2$ GeV, $\sin^2 2\theta_{\mu\tau}  \simeq 0.87$,
$n_L/s \simeq -1.8 \times 10^{-10}$ and $\theta \simeq 0.016$ for
$\delta \simeq -\pi/3$.  
\vskip 0.25cm

It is interesting to note that the mass scale $v_0$ is of order
$10^{16}$ GeV which is consistent
with the unification of the gauge couplings of the MSSM. Also, the values
of the
$\mu$-neutrino mass, for which solutions are found, turn out to be
consistent with the large rather than the small angle MSW mechanism.
\vskip 0.25cm

In summary, we found that, within the framework of our inflationary scheme, 
a number of cosmological and phenomenological constraints could be easily 
satisfied. These constraints include the gravitino bound on the reheat 
temperature, successful baryogenesis in the universe, and consistency of 
neutrino masses and mixing with the solar and atmospheric neutrino 
oscillations. It should be pointed out that, for this success of our model, 
no extra structure was needed to be added to it. Actually, the possibility 
of accounting for these requirements is more or less automatic
in this inflationary scheme and can, in principle, remain in gauge 
groups other than $G_{PS}$ too.

\section{ Conclusions}

We have constructed a SUSY GUT model based on the $G_{PS}$ gauge symmetry 
group. This model is consistent with all particle physics and cosmological
requirements. The $\mu$-problem is solved by introducing a global
anomalous  PQ symmetry $U(1)_{PQ}$, which also solves the strong CP
problem. Although baryon and lepton numbers are violated in the
superpotential, the proton turns out to be practically stable. SUSY hybrid
inflation is `naturally'  and successfully incorporated in this model but
in an unconventional way.  In the standard realizations of SUSY hybrid
inflation, the superpotential involves only renormalizable couplings of
the GUT Higgs superfields and a gauge singlet. We have modified this
picture by including the next order non-renormalizable coupling too. In
contrast to the usual case, inflation now takes place along a classically
flat direction where the gauge symmetry  ($G_{PS}$) is spontaneously
broken to $G_{SM}$. As a consequence, after inflation ends, there is
absolutely no production of doubly charged magnetic monopoles, which are
associated with the breaking of $G_{PS}$. Thus, the cosmological
catastrophe
one would encounter by employing the usual inflationary scheme in the
SUSY PS theory is avoided. Our mechanism is crucial for the viability of
any model containing cosmologically disastrous topological defects such as
magnetic monopoles or domain walls and leads to complete absence of such
objects. It is interesting to point out that, although the usual
trajectory with unbroken $G_{PS}$ also exists, there is a range of
parameters for which the system finally inflates along the non-trivial
path before falling into the SUSY vacua. Thus, the monopole problem can
be solved for all possible initial conditions. 
\vskip 0.25cm
   The classical flatness of the inflationary valley is lifted by 
one-loop radiative corrections which produce an inclination for driving 
the inflaton towards the SUSY vacua. The measurements of COBE can be 
easily reproduced with natural values (of order $10^{-3}$) of the 
relevant coupling constant. The GUT mass scale comes out equal to (or 
somewhat smaller than) the SUSY GUT scale and can certainly be much 
closer to it than in the standard SUSY inflationary scheme. The spectral 
index of density perturbations ranges between about 1 and 0.9.
\vskip 0.25cm
After inflation ends, the inflaton oscillates about 
the SUSY vacuum and decays into the second heaviest right-handed 
neutrino superfield thereby reheating the universe. The subsequent decay 
of these right-handed neutrinos to lepton and electroweak Higgs 
superfields generates a lepton asymmetry which is then 
partially converted to baryon asymmetry by the electroweak instantons.
We require that the so obtained baryon asymmetry of the universe is
consistent with the low deuterium abundance constraint. We also take
almost maximal $\nu_{\mu}-\nu_{\tau}$ mixing as indicated by
SuperKamiokande. The $\mu$- and $\tau$-neutrino masses are restricted by
the MSW resolution of the solar neutrino puzzle and the heaviest Dirac
neutrino mass by
$SU(4)_c$ symmetry. We find that all these requirements can be met with
natural values (of order $10^{-3}$) of the relevant coupling constant.
Note that the second heaviest Dirac neutrino mass
turns out to be of order $1$ GeV and masses of $\nu_{\mu}$ consistent
with the large rather than the small angle MSW mechanism are favored.
\vskip 0.25cm
   
Finally, we would like to point out that, although we restricted our 
discussion to the PS gauge group, this new SUSY hybrid inflationary 
scenario is of much wider applicability. The reason for choosing this 
particular framework for our presentation is that $G_{PS}$ is one 
of the simplest gauge groups with magnetic monopoles. The scenario 
can be readily extended to other semi-simple gauge groups such as 
the trinification group $SU(3)_c\times SU(3)_L\times SU(3)_R$ 
which emerges from string theory. Of course, the superfields $H^c$ 
and $\bar{H}^c$ should be replaced by the appropriate pair of 
superfields $(1,\bar 3,3)$ and $(1,3,\bar 3)$ from the $E_6$ 
$27$-plet and $\overline{27}$-plet which cause the breaking of the 
trinification group to $G_{LR}$. Further breaking of $G_{LR}$ 
to $G_{SM}$ is achieved by a similar pair of superfields with 
vevs, though, in the right-handed neutrino direction. Our scheme 
could also be extended to simple gauge groups such as $SO(10)$. 
The breaking of $SO(10)$ can be achieved by including, among other 
representations, a pair of $16$, $\overline{16}$ Higgs superfields 
acquiring vevs in the right-handed neutrino direction. Inflation 
and reheating are expected to be quite similar to the ones 
discussed here with the only complication that more Higgs superfield 
representations such as $54$ and $45$ will be involved for gauge 
symmetry breaking to $G_{SM}$. The  magnetic monopole problem can 
then be solved only if some of these fields are non-zero too on the 
inflationary trajectory.

\section*{Acknowledgement}
   This work was supported by European Union under the TMR contract
No. ERBFMRX-CT96-0090. S. K. is supported by a Ministerio de Educacion y
Cultura research grant. Also, Q. S. would like to acknowledge the DOE 
support under grant number DE-FG02-91ER40626. Finally, R. J. and  S. K. 
would like to thank G. Senjanovi\'{c} for discussions.

%\newpage
\section*{\hspace{-.07cm}Appendix: Derivation of the mass 
spectrum during\\
inflation}
 In this Appendix, we sketch the derivation of the mass spectrum
of the model when the system is trapped along the inflationary
trajectory at $y_2$. During inflation, the fields $H^c$,
$\bar{H}^c$ acquire vevs in the $\nu^c_H$, $\bar{\nu}^c_H$
direction which break the gauge symmetry $G_{PS}$ down to
$G_{SM}$. These vevs are given by $\langle \nu^c_H \rangle =
\langle \bar{\nu}^c_H \rangle = v =(\kappa M_S^2/2\beta)^{1/2}$
and we can write $\nu^c_H = v + \delta \nu^c_H$ and $\bar{\nu}^c_H
= v + \delta \bar{\nu}^c_H$. One can show that the scalar
potential in Eq.(\ref{eq:infpot}) does not generate masses for the scalar
components of the superfield $H^c$ $(\bar{H}^c)$ in the directions
$u^c_H$, $d^c_H$, $e^c_H$ ($\bar{u}^c_H$, $\bar{d}^c_H$,
$\bar{e}^c_H$). On the contrary, a simple calculation yields that
the normalized real scalar fields $\mathrm{Re}(\delta \nu^c_H +
\delta \bar{\nu}^c_H)$ and $\mathrm{Im}(\delta \nu^c_H + \delta
\bar{\nu}^c_H)$ acquire non-zero masses given by
\begin{equation}
m_{\pm}^2=4\kappa^2 \vert S \vert^2 \mp 2\kappa^2M^2(\frac{1}{4
\xi}-1) \label{a1}
\end{equation}
respectively. The superpotential in Eq.(\ref{eq:susyinfl}) gives rise 
to just one
massive Majorana fermion with $m^2=4 \kappa^2 \vert S \vert^2$
corresponding to the direction $(\nu^c_H +
\bar{\nu}^c_H)/\sqrt{2}$. We see that the SUSY breaking
along the inflationary trajectory, which is due to the non-zero
vacuum energy density $\kappa^2 M^4 (1/4\xi-1)^2$, produces a mass
splitting in the $\nu^c_H$, $\bar{\nu}^c_H$ supermultiplets.
Actually, as we will show, this is the only place where such a mass
splitting appears. \vskip 0.25cm

 The D-term contribution to the scalar masses can be found from
\begin{equation}
\frac{1}{2} g^2 \sum_{a}(\bar{H}^{c^*} T^a \bar{H}^c + H^{c^*} T^a
H^c)^2, \label{a2}
\end{equation}
where $g$ is the $G_{PS}$ gauge coupling constant and the sum
extends over all the generators $T^a$ of $G_{PS}$. The part of
this sum over the generators $T^{15}=(1/2\sqrt{6})\
\mathrm{diag}(1, 1, 1, -3)$ of $SU(4)_c$ and
$T^3=(1/2)\ \mathrm{diag}(1, -1)$ of $SU(2)_R$ gives rise to a mass
term for the normalized real scalar field $\mathrm{Re}(\delta
\nu^c_H - \delta \bar{\nu}^c_H)$ with $m^2=5g^2v^2/2$ as one can
show by using the above expansion of $\nu^c_H$, $\bar{\nu}^c_H$.
\vskip 0.25cm
   The gauge bosons $A^a$ can acquire masses from the Lagrangian
terms
\begin{equation}
g^2( \vert \sum_{a} A^a T^a \bar{H}^c \vert^2+ \vert \sum_{a}A^a
T^a H^c \vert^2). \label{a3} \end{equation} Taking the
contribution of $T^{15}$ and $T^3$ again, we obtain a mass term
for the normalized gauge field $$A^{\perp} =-\sqrt{\frac{3}{5}}
A^{15} + \sqrt{\frac{2}{5}}A^3$$ with $m^2=5g^2v^2/2$ (the real
field $\mathrm{Im}(\delta \nu^c_H - \delta \bar{\nu}^c_H)$, which
is so far left massless, is absorbed by this gauge boson). \vskip
0.25cm

   Fermion masses get also contributions from the Lagrangian terms
\begin{equation}
i \sqrt{2} g \sum_{a} \lambda^a (\bar{H}^{c^*} T^a
\psi_{\bar{H}^c}+ H^{c^*} T^a \psi_{H^c})+h.c., \label{a4}
\end{equation}
where $\lambda^a$ is the gaugino corresponding to $T^a$ and
$\psi_{\bar{H}^c}$, $\psi_{H^c}$ represent the chiral fermions
belonging to the superfields $\bar{H}^c$, $H^c$ respectively.
Concentrating again on $T^{15}$, $T^3$, we obtain a Dirac mass
term between the chiral fermion in the superfield $(\nu^c_H -
\bar{\nu}^c_H)/\sqrt{2}$ and the gaugino $-i \lambda^{\perp}$ with
$m^2=5g^2v^2/2$. This completes the analysis of the $\nu^c_H$,
$\bar{\nu}^c_H$ sector together with the gauge supermultiplet in the
$T^{\perp}$ direction.
\vskip 0.25cm

The eight normalized real scalar fields $\mathrm{Re}(u^c_H -
\bar{u}^{c^*}_H)$, $\mathrm{Im}(u^c_H -\bar{u}^{c^*}_H)$ 
(three colors), $\mathrm{Re}(e^c_H -
\bar{e}^{c^*}_H)$, $\mathrm{Im}(e^c_H - \bar{e}^{c^*}_H)$ acquire
mass terms from the D-term contribution in Eq.(\ref{a2}) with
$m^2=g^2v^2$. Indeed, the part of the sum in this equation over
the generators
\begin{equation}
T^1=\frac{1}{2} \left(\begin{array}{cc}
        0 & 1\\
        1 & 0 \end{array} \right), \hspace{1.5cm}
T^2=\frac{1}{2} \left(\begin{array}{cc}
                0 & -i\\
                i & 0 \end{array} \right)
\label{a5}
\end{equation}
of $SU(2)_R$ gives
\begin{equation}
\frac{1}{2} g^2 v^2 (\mathrm{Re}(e^c_H - \bar{e}^{c^*}_H))^2 +
\frac{1}{2} g^2 v^2 (\mathrm{Im}(e^c_H - \bar{e}^{c^*}_H))^2.
\label{a6}
\end{equation}
Similarly, the sum over the $SU(4)_c$ generators $T^1_i$ and
$T^2_i$ $(i=1,2,3)$ with $1/2$ $(1/2)$ and $-i/2$ $(i/2)$ in the
$i4$ $(4i)$ entries respectively and zero everywhere else
generates the masses of $\mathrm{Re}(u^c_H - \bar{u}^{c^*}_H)$, 
$\mathrm{Im}(u^c_H - \bar{u}^{c^*}_H)$ (three colors). Using
Eq.(\ref{a3}), one can show that the eight gauge bosons $A^1$,
$A^2$, $A^1_i$, $A^2_i$ $(i=1,2,3)$ become massive with
$m^2=g^2v^2$ (they absorb the real fields $\mathrm{Re}(e^c_H +
\bar{e}^{c^*}_H)$, $\mathrm{Im}(e^c_H + \bar{e}^{c^*}_H)$, 
and $\mathrm{Re}(u^c_H + \bar{u}^{c^*}_H)$,
$\mathrm{Im}(u^c_H + \bar{u}^{c^*}_H)$ (three colors)). \vskip
0.25cm

   The chiral fermions $\psi_{\bar{e}^c_H}$ and  $\psi_{e^c_H}$ 
combine with
the gauginos $\lambda^+ = (\lambda^1 + i \lambda^2)/\sqrt{2}$ and
$\lambda^- = (\lambda^1 - i\lambda^2)/\sqrt{2}$ respectively to
form two Dirac fermion states with $m^2=g^2v^2$ as one deduces
from Eq.(\ref{a4}). Similarly, $\psi_{\bar{u}^c_H}$ and
$\psi_{u^c_H}$ (three colors) together with $\lambda^+_i=
(\lambda^1_i + i\lambda^2_i)/\sqrt{2}$ and $\lambda^-_i=
(\lambda^1_i - i\lambda^2_i)/\sqrt{2}$ give six more Dirac
fermions with the same mass squared. \vskip 0.25cm

   The only superfields left are the $d^c_H$, $\bar{d}^c_H$, $g^c$, 
$\bar{g}^c$. They
do not mix with the rest of the spectrum, as one can easily show,
and acquire masses from the last two superpotential terms in
Eq.(\ref{eq:superpot}). These terms can be explicitly written as
\begin{eqnarray}
a\ G\ H^c\ H^c\ &=& 2\ a\ (-d^c_H\ \nu^c_H\ +\ u^c_H\ e^c_H)\
\bar{g}^c\ + 2\ a\ u^c_H\ d^c_H\ g^c,\nonumber\\ b\ G\ \bar{H}^c\
\bar{H}^c &=& 2\ b\ (-\bar{d}^c_H\ \bar{\nu}^c_H\ +\ \bar{u}^c_H\
\bar{e}^c_H)\ g^c + 2\ b\ \bar{u}^c_H \bar{d}^c_H \bar{g}^c.
\label{a7}
\end{eqnarray}
The scalar potential then contains the terms
\begin{equation}
4 a^2\ v^2\ (\vert d^c_H \vert^2+ \vert \bar{g}^c \vert^2)\ +\ 4\
b^2\ v^2\ (\vert \bar{d}^c_H \vert^2+ \vert g^c \vert^2)
\label{a8}
\end{equation}
and we obtain six complex scalars $(d^c_H, \bar{g}^c)$ with
$m^2=4a^2v^2$ and six complex scalars $(\bar{d}^c_H, g^c)$ with
$m^2=4b^2v^2$. Also, the chiral fermions $\psi_{d^c_H}$ and
$\psi_{\bar{g}^c}$ combine to give three Dirac fermions with
$m^2=4a^2v^2$, while $\psi_{\bar{d}^c_H}$ and $\psi_{g^c}$ give
three Dirac fermions with $m^2=4b^2v^2$. \vskip 0.25cm

   We see that all the fields acquire non-zero masses but SUSY
is broken only in the sector of the two real scalar fields with
masses given in Eq.(\ref{a1}) and the Majorana fermion with
$m^2=4\kappa^2 \vert S \vert^2$. In all other supermultiplets, the
fermionic and bosonic components have equal masses. As a
consequence, these supermultiplets give zero contribution to the
supertraces $\mathrm{STr}M^{2n}$ for any integer $n \geq 0$ (and
in general to the supertrace of any function of the mass squared
matrix $M^2$). Thus, in calculating any such supertrace, we only
have to consider the two real scalars and the Majorana fermion
mentioned above. Note that, in particular, their contribution to
$\mathrm{STr}M^2$ is zero.

%\newpage

\newpage
\section*{Figures}
\vskip 0.85cm
\begin{figure}[ht]
\psfig{figure=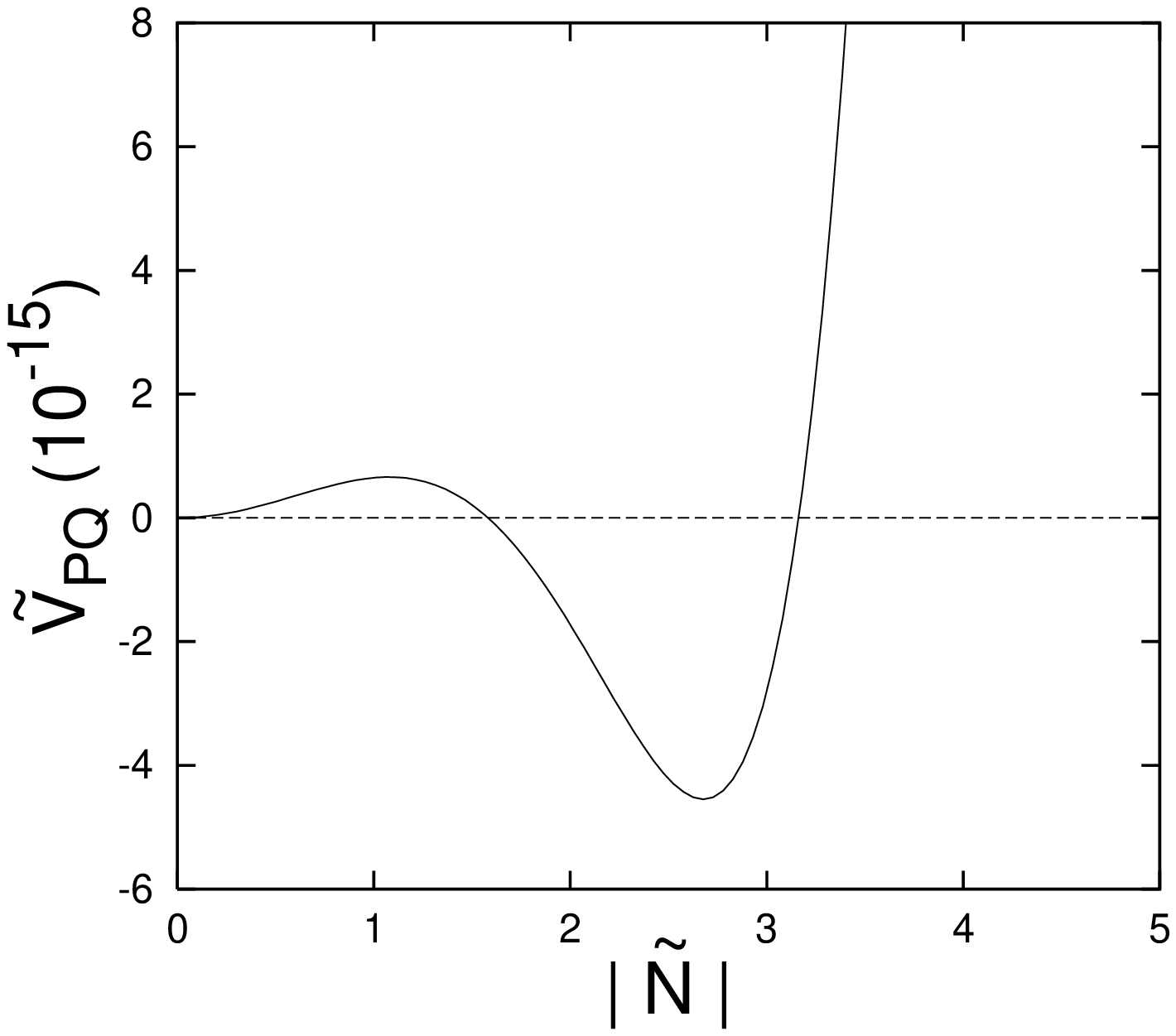,height=8cm,width=13cm} \caption{ The
(dimensionless) zero temperature potential $\tilde{V}_{PQ}=V_{PQ}/(m_{3/2}
M_S)^2$ with $V_{PQ}$ given in Eq.(\ref{potn0}) versus $\vert \tilde{N}
\vert=\vert N \vert/(m_{3/2} M_S)^{1/2}$, for $\vert A \vert=5,
\lambda_1=0.3, \lambda_2=0.1, m_{3/2}=300$ GeV and $M_S=5\times
10^{17}$ GeV ($\mu = 600$ GeV). } \label{vn0} \vskip 0.4cm
\end{figure}

\begin{figure}[ht]
\psfig{figure=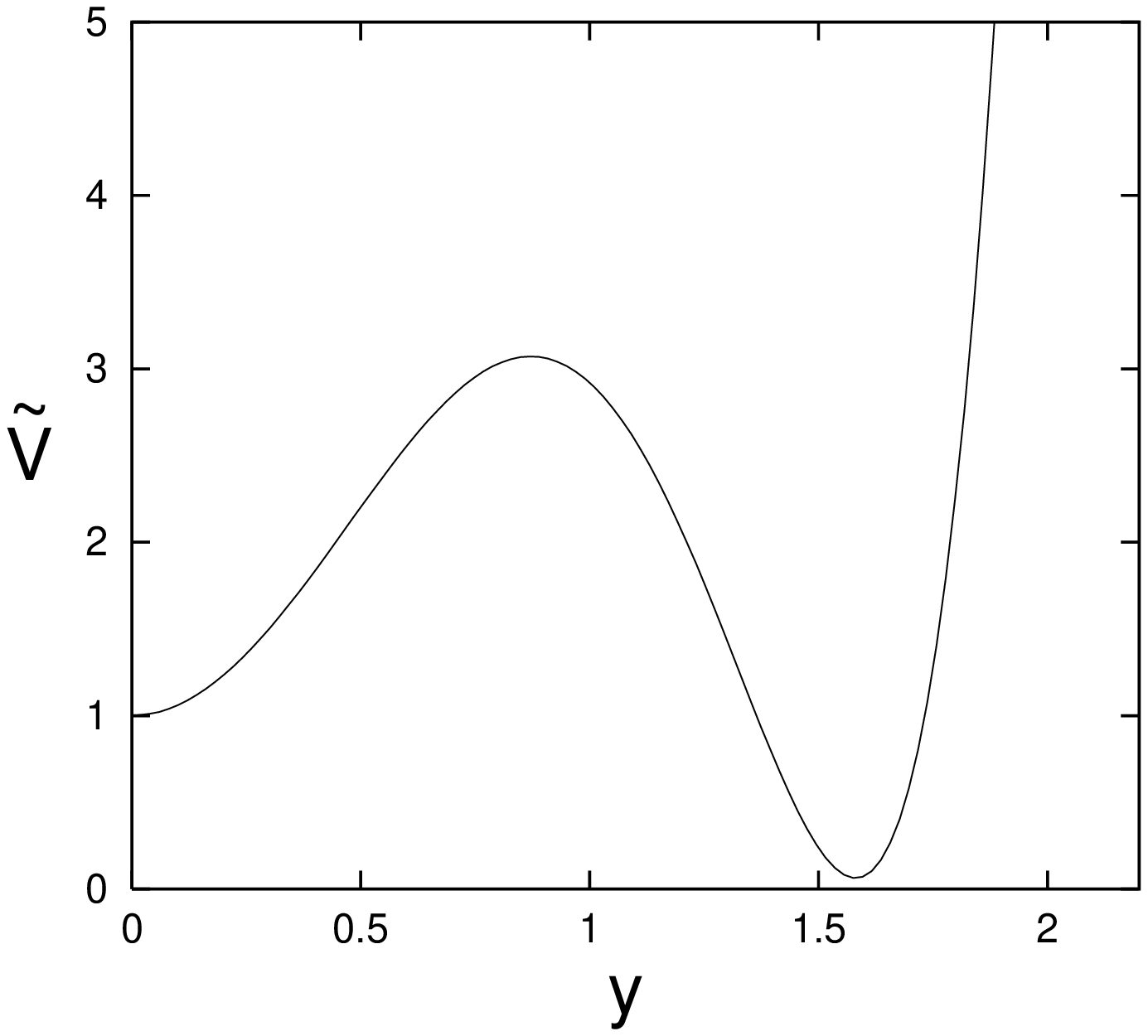,height=8cm,width=13cm} \caption{The
(dimensionless) potential
$\tilde{V}$, given in Eq.(\ref{vtilde}), versus $y$ for $w=2$, 
$\xi=1/5$.}\vskip 0.5cm
\end{figure}

\begin{figure}[ht]
\psfig{figure=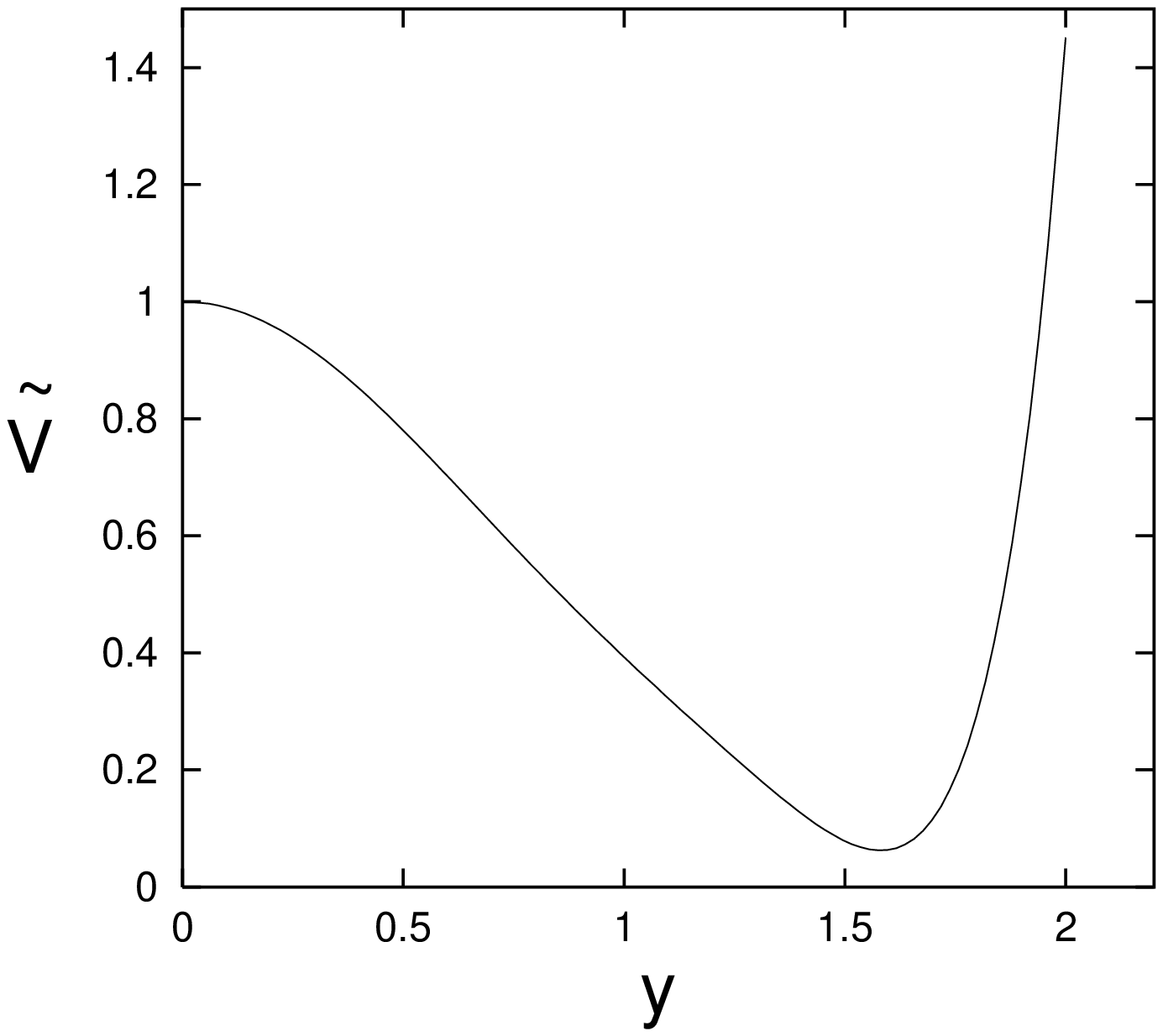,height=8cm,width=13cm} \caption{The
(dimensionless) potential
$\tilde{V}$, given in Eq.(\ref{vtilde}), versus $y$ for $w=0.7$, 
$\xi=1/5$.}\vskip 0.5cm
\end{figure}

\begin{figure}[ht]
\psfig{figure=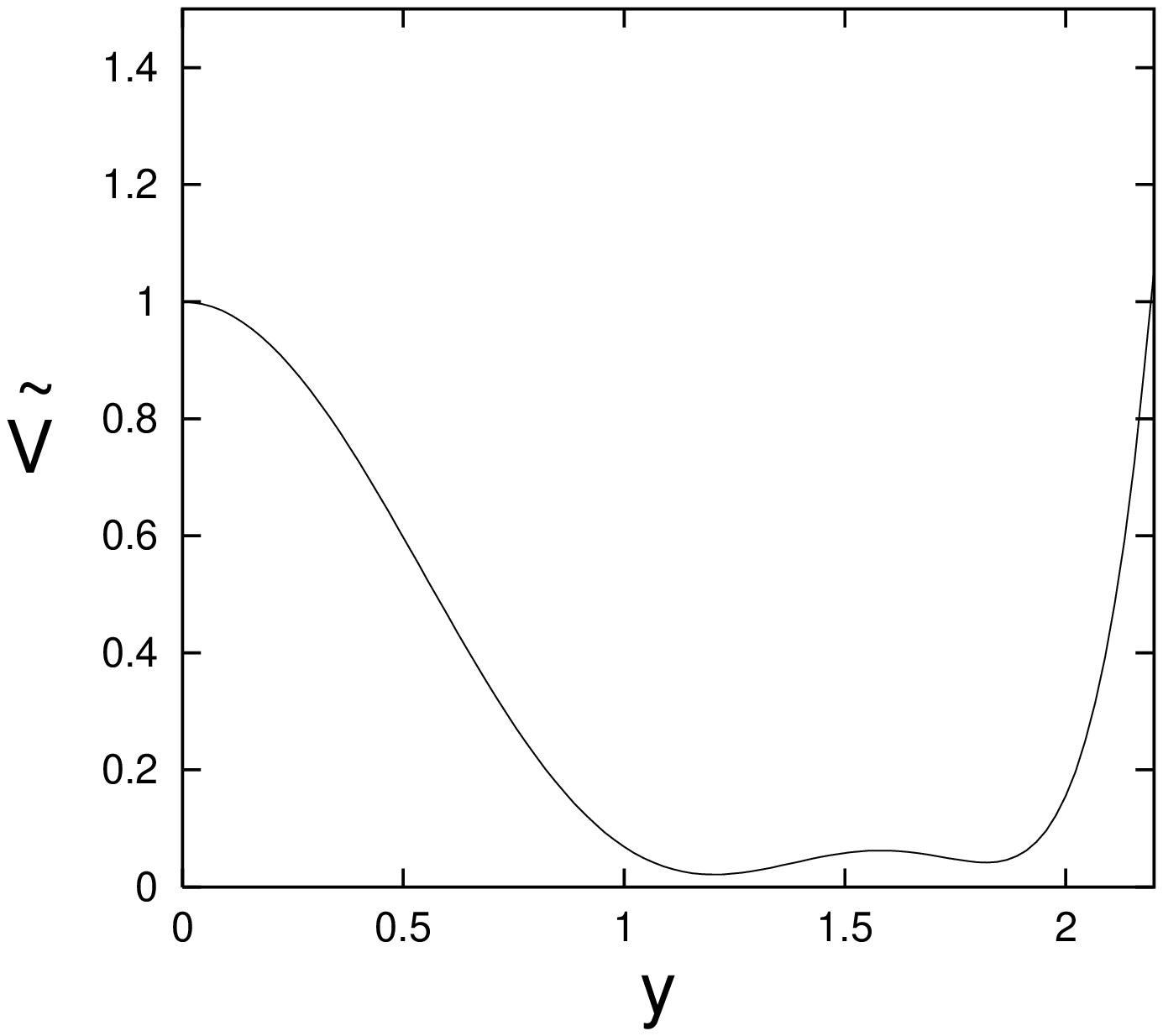,height=8cm,width=13cm} \caption{The
(dimensionless) potential
$\tilde{V}$, given in Eq.(\ref{vtilde}), versus $y$ for $w=0.2$, 
$\xi=1/5$.}\vskip 0.5cm
\end{figure}

\begin{figure}[ht]
\psfig{figure=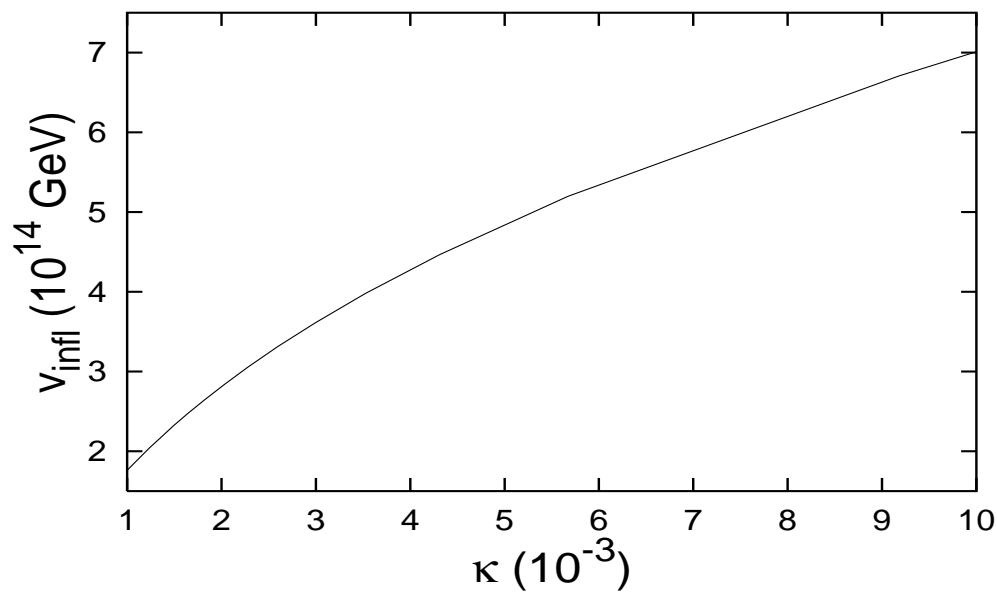,height=8cm,width=13cm} \caption{The
inflationary scale $v_{\mathrm{infl}}$ as a function of the coupling
constant
$\kappa$.} \label{fig:fig1} \vskip 1.25cm
\end{figure}

\begin{figure}[ht]
\psfig{figure=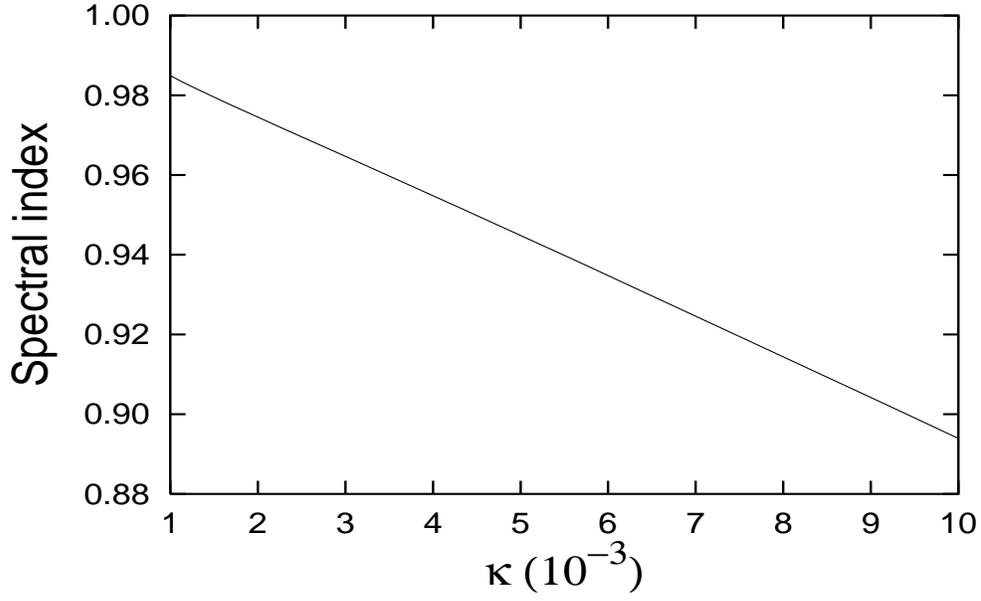,height=8cm,width=13cm} \caption{The
spectral index $n$ as a function of the coupling constant $\kappa$.}
\label{fig:fig2}\vskip 0.5cm
\end{figure}

\begin{figure}[ht]
\psfig{figure=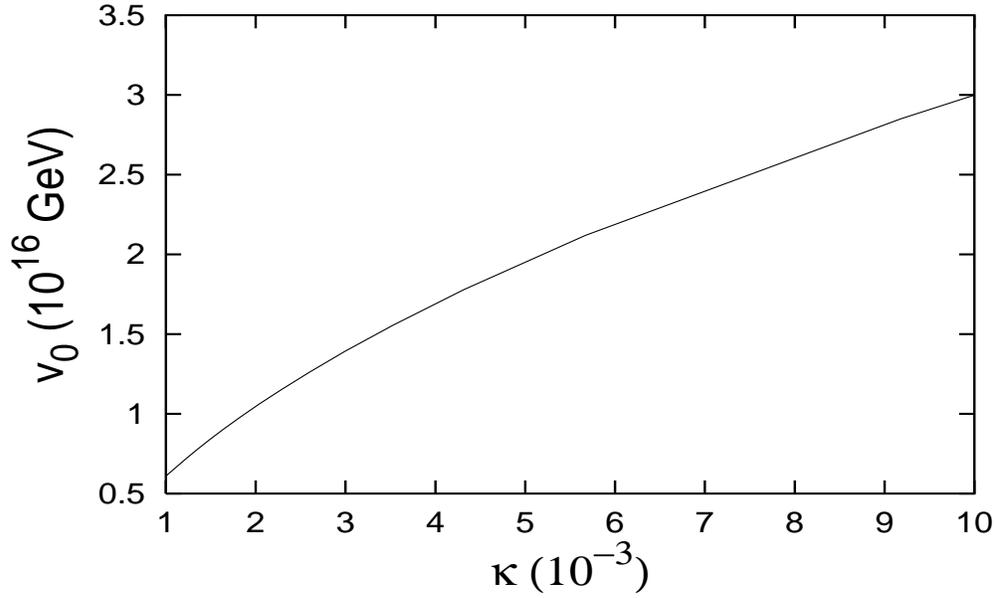,height=8cm,width=13cm} \caption{The common
vev $v_0$ of $\bar{H}^c$ , $H^c$ at the SUSY
minimum as a function of the coupling constant $\kappa$.} 
\label{fig:fig3}
\end{figure}

\end{document}